# Device-Edge Cooperative Fine-Tuning of Foundation Models as a 6G Service

*Hai Wu, Xu Chen, and Kaibin Huang*[1]

## Abstract

Foundation models (FoMos), referring to large-scale AI models, possess human-like capabilities and are able to perform competitively in the domain of human intelligence. The breakthrough in FoMos has inspired researchers to deploy such models in the sixth-generation (6G) mobile networks for automating a broad range of tasks in next-generation mobile applications. While the sizes of FoMos are reaching their peaks, their next phase is expected to focus on fine-tuning the models to specific downstream tasks. This inspires us to propose the vision of FoMo fine-tuning as a 6G service. Its key feature is the exploitation of existing parameter-efficient fine-tuning (PEFT) techniques to tweak only a small fraction of model weights for a FoMo to become customized for a specific task. To materialize the said vision, we survey the state-of-the-art PEFT and then present a novel device-edge fine-tuning (DEFT) framework for providing efficient and privacy-preserving fine-tuning services at the 6G network edge. The framework consists of the following comprehensive set of techniques: 1) Control of fine-tuning parameter sizes in different transformer blocks of a FoMo; 2) Over-the-air computation for realizing neural connections in DEFT; 3) Federated DEFT in a multi-device system by downloading a FoMo emulator or gradients; 4) On-the-fly prompt-ensemble tuning; 5) Device-to-device prompt transfer among devices. Experiments are conducted using pre-trained FoMos with up to 11 billion parameters to demonstrate the effectiveness of DEFT techniques. The article is concluded by presenting future research opportunities.

## 1. Introduction

Representing the latest breakthrough in Artificial Intelligence (AI), large-scale Foundation Models (FoMos) are capable of performing tasks in the domain of human intelligence, including producing original songs, poems, essays, digital photos, and drawings at professional levels. In particular, ChatGPT, a prevalent FoMo focusing on natural language processing, has excelled in career exams designed for people, e.g., performing in the 90th percentile in the American Uniform Bar Examination. Furthermore, multi-modal FoMos incorporating sensing, speech and vision are under active development to control robots and navigate drones and vehicles. The emergence of FoMos' capabilities is the joint effect of four factors: 1) an enormous model size (e.g., 175 billion parameters for GPT-3), 2) billions of self-supervised training runs, 3) astonishing quantities of high-quality unlabelled data collected across the entire Internet, and 4) an attention network to efficiently learn the relations between words and concepts [1, 2]. The promise of FoMos attracts tech giants like Alphabet, Amazon, and Nvidia to invest heavily in training their own FoMos for incorporation into their services. In the area of mobile networks, researchers are also inspired to deploy FoMos in the sixth-generation (6G) mobile networks to automate tasks of mobile devices such as human semantic communications, personal assistants, auto-pilot, and robotic control [3].

---

[1] The authors are with Dept. of Electrical and Electronic Engineering, The University of Hong Kong, Hong Kong. Contact: K. Huang (email: huangkb@eee.hku.hk).

FoMos pre-trained on generic data are required to be fine-tuned based on specific tasks to maximize the models' effectiveness, e.g., personalized conversation by learning a user's background, personality, and life habits. Training a FoMo to a specific task, however, requires enormous computation resources and storage space. For example, 1.3 gigawatt-hours of electricity and USD 4.6 million were spent on training ChatGPT-3. A more practical approach is called parameter-efficient fine-tuning (PEFT), referring to a class of techniques that vary only a small fraction of model weights or architecture to achieve maximum task effectiveness. A recent study by OpenAI, the company that created ChatGPT, reveals that fine-tuned FoMos are capable of undertaking 19,000 tasks involving 1,016 occupations. Research on PEFT leads to a fast-growing area with rich literature (§ 2).

In mobile networks, on-device execution of FoMos is impractical but they can be deployed at edge servers to allow low-latency remote access. This is closely aligned with one key mission of 6G to realize an intelligent network edge featuring ubiquitous AI and learning as a platform for supporting next-generation mobile applications [4]. In return, FoMos at the edge can also benefit from real-time mobile data in proximity. This helps to overcome the limit of training data that stymies their continued improvement as the data in the public domain have been exhausted. FoMos at the 6G edge is embraced as a platform for automating a broad range of IoT applications such as elderly care, personal assistant, robotic control, auto pilot, and augmented reality. To materialize the vision, it is proposed in this article that FoMo fine-tuning is provided as a 6G service (§ 3). Wireless communication in 6G is shifting from generic rate-centric designs towards goal-oriented communications designed using a communication-computing integrated approach. Aligned with this paradigm shift, we further design a novel framework called device-edge fine tuning (DEFT) to provide goal-oriented wireless techniques for supporting efficient, privacy-preserving device-edge cooperation in fine-tuning FoMos. The framework comprises a set of novel DEFT techniques summarized as follows.

- The control of fine-tuning parameter size in different transformer blocks of a FoMo to cope with devices' resources constraints in DEFT (§ 4.1);

- Over-the-air computation for realizing neural connections in DEFT with a split FoMo (§ 4.2);

- Federated DEFT in a multi-device system via downloading of either a FoMo emulator or gradients (§ 5.1);

- On-the-fly prompt-ensemble tuning based on a boosting approach (§ 5.2);

- Device-to-Device (D2D) prompt transfer among devices with similar tasks (§ 5.3).

We have also conducted experiments of fining-tuning pre-trained FoMos, one called Megatron-11B, which consists of around 11 billion parameters, and another one is T5-Base with 200 million parameters, on a Nvidia A100 server to demonstrate the effectiveness of DEFT (§ 6).

## 2. State-of-the-Art Techniques for PEFT

The prevalent Transformer architecture adopted for FoMo pre-training is shown in Fig. 1, which can be fine-tuned using one or multiple of the following PEFT techniques.

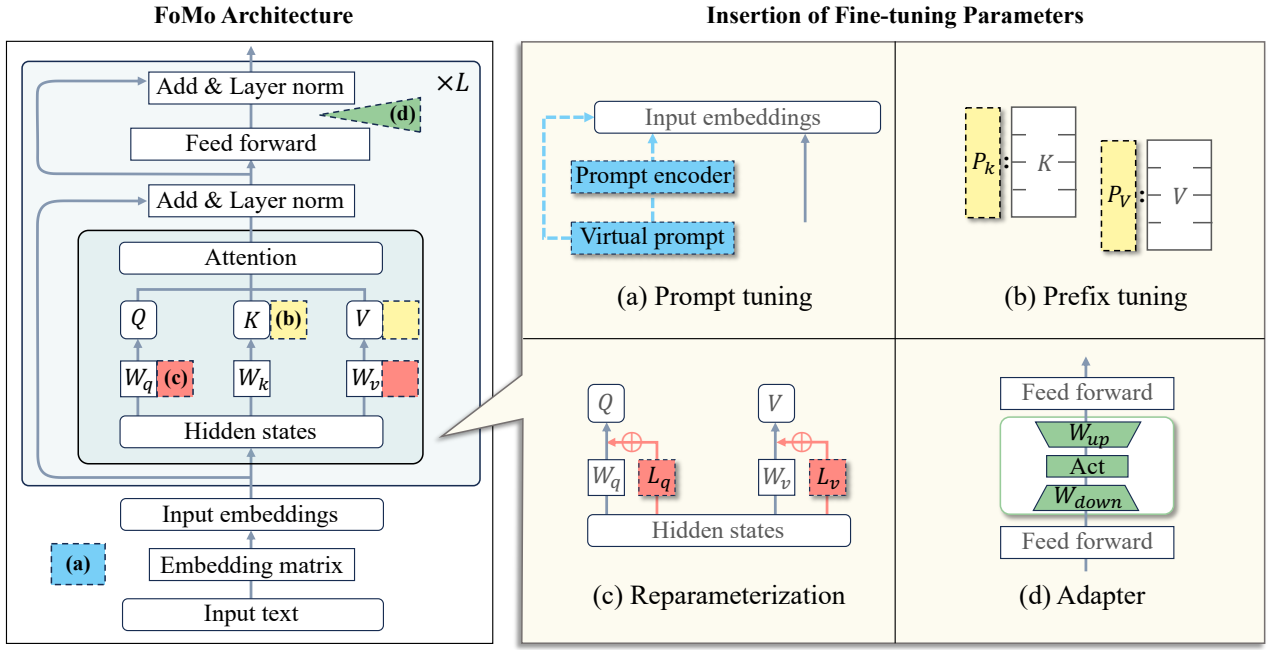

Fig. 1. State-of-the-Art PEFT techniques.

## 2.1 Soft Prompts

A prompt refers to input information that can help a FoMo understand the current context or task so as to interact with or assist a user more effectively [1]. In contrast with a handcrafted prompt, a soft prompt is a variable matrix representing the tunable parameters. It is either input into a pre-trained FoMo together with input embeddings or appended to attention matrices in each Transformer block of the FoMo, corresponding to two techniques, P-tuning [see Fig. 1(a)] and Prefix-tuning [see Fig. 1(b)], respectively.

*1) P-tuning:* A soft prompt is generated using a lightweight encoder, typically a bidirectional long-short memory network (LSTM) with two successive multilayer perceptron (MLP), with a chosen initial prompt as the input [5]. LSTM is used because it can associate different elements ("tokens") in the soft prompt. Then, the fine-tuning process is to train the encoder for a specific task via back-propagated gradient descent. Its completion yields the desired soft prompt that is attached to every instance of input embeddings in subsequent task execution while the encoder can be discarded.

*2) Prefix-tuning:* Both the key and value attention matrices in each FoMo transformer block are attached with prefix matrices (i.e., soft prompts) [6]. The prefixes are then tuned using an MLP for a given task to overcome the instability faced in direct optimization. Prefix-tuning is demonstrated to achieve comparable text-generating performance as full-model fine-tuning by optimizing only around 1% parameters. Its performance is highly dependent on the prefix length.

## 2.2 Reparameterization Methods

Fine-tuning a large language model towards a downstream NLP task is demonstrated to be performed only in an extremely low-dimensional parametric subspace [7]. Inspired by this fact, the essential idea of the reparameterization-based techniques is to approximate the weight matrix of each block of a fine-tuned FoMo by the superposition of the original counterpart with a tunable extremely low-rank matrix, thereby reducing the computation overhead of fine-tuning. The principle can be implemented by adding an additional data path with low-rank weight matrices to each FoMo block as illustrated in Fig. 1(c). There exist three common methods for constructing a low-rank matrix, called a fine-tuning matrix: *1) Fastfood transformation:* a fine-tuning matrix is the product of several pre-defined, randomly generated matrices and a low-dimensional tunable matrix. *2) Low-rank approximation:* a fine-tuning matrix is given by the product of two low-rank matrices with preset ranks. *3) Kronecker product:* a fine-tuning matrix is given by the Kronecker product of several low-dimensional components.

## 2.3 Adapters and Selective Parameters

An adapter refers to a lightweight, tunable neural network consisting of only several layers. Then, an adapter-based PEFT technique inserts an adapter into the original FoMo architecture and tunes the adapter to a specific task by applying the back-propagated gradient descent to the modified architecture. The originally proposed adapter comprises two feedforward projection layers sandwiching a nonlinearity activation layer [8]. The initial design has been substantially generalized to have a variable number of layers of different types and a flexible location in the backbone neural network. Instead of inserting an adapter, a selective-parameter technique is simpler as it fine-tunes a small subset of layers (or a small subset of parameters) of the original FoMo while freezing the rest. For example, the fine-tuned parameters could be the first few layers of a FoMo, the model's bias parameters, or specific rows of model weights [9]. Alternatively, a selective-parameter technique can make an unstructured selection of parameters in FoMos for sparse model updating. There exist a number of selection criteria such as the top-K rule or one based on optimizing a sparse binary mask for weight selection.

## 2.4 Hybrid Methods

The PEFT techniques discussed are complementary to each other. Hence, combining multiple techniques (e.g., sparse adapter [9]) can further improve the fine-tuning performance.

# 3. PEFT as a 6G Service

The advancement in general-purpose FoMo is believed to reaching its peak. The next phase of FoMo aims at developing task-oriented FoMos via fine-tuning or highly compact mobile FoMos. For instance, OpenAI offers an interface with ChatGPT to allow customers to fine-tune the FoMo according to their needs; Qualcomm demonstrates a sub-10 billion parameter FoMo for mobile devices; Google is developing a multi-modal FoMo targeting robotics. In view of this above trend and the broad range of IoT applications 6G aims to support, we propose that PEFT be provided as a service to mobile users to automate relevant tasks. As mentioned, FoMos can also benefit from their integration with mobile networks to overcome the data bottleneck and attain continued improvements. The PEFT services can be implemented in different ways. One is to store a large library of task-specific and thus highly compact FoMos in the central and edge clouds for the user to download according to their real-time demands [10]. This approach is limited by the availability of task-specific data to the cloud as a large portion is private data. The alternative implementation is the DEFT framework.

The basic principle of DEFT is to coordinate on-device and on-server computing to optimize fine-tuning parameters using local data and a FoMo operated by a server (or servers) based on stochastic gradient descent, as shown in Fig. 2. One approach is split learning that splits a FoMo with fine-tuning

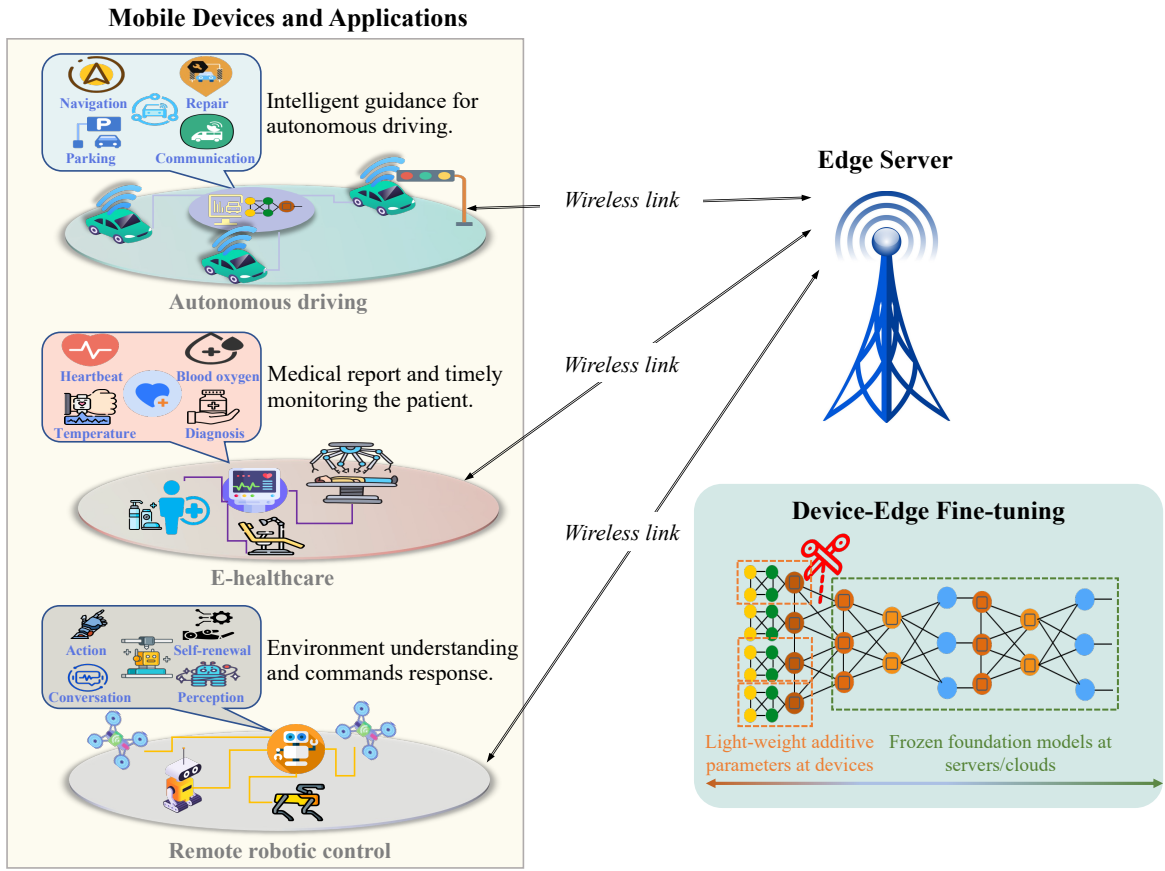

Fig. 2. Device-edge cooperative fine-tuning in 6G networks and relevant applications.

parameters into device and server sub-models, which are conducted by air interface. This not only allows fine-tuning to leverage mobile computation resources but also helps to preserve user privacy by avoiding raw data uploading. Ideally, fine-tuning parameters should be kept on-device and updated using local data. For instance, considering the selective-layer PEFT techniques, the fine-tuned layers can be executed on a device with the remaining FoMo network executed by the server. Other DEFT approaches include uploading data embeddings and downloading a low-dimensional FoMo emulator. Several key challenges faced by PEFT implementation are described below and related to

- Mobile resource constraint: The performance of fine-tuned FoMo improves as the number of fine-tuning parameters and/or computation time and/or the amount of local data grows. However, the improvement is limited by the mobile constraints on computation resources and storage space. To overcome the limitation benefits from increasingly powerful mobile AI processors. The issue can also be addressed from the perspective of algorithmic design (e.g., see § 4.1).

- User privacy preservation: A naive approach is to employ an existing PEFT technique at an edge server to fine-tune a FoMo. However, the required uploading of users 'personal data would compromise their ownership. Therefore, data privacy preservation is an important factor to consider in designing DEFT (e.g., see § 5.1).

- Goal-oriented air interface: Each iteration in the DEFT process involves the device and server exchanging high-dimensional data, i.e., data embeddings, prompt, gradient, or parameters. The resultant communication bottleneck is exacerbated in the scenario of multi-device cooperation. This calls for the design of a goal-oriented air interface for DEFT, which integrates computing and communication to achieve high communication efficiency. We propose relevant techniques in §§ 4.2, 5.2, and 5.3.

# 4. Single-User DEFT

## 4.1 Fine-tuning Parameter Allocation for DEFT

A PEFT technique (e.g., reparameterization methods) typically adds fine-tuning parameters to individual (transformer) blocks of a FoMo. In the context of DEFT, the communication overhead tends to increase rapidly as the parameter size grows. On the other hand, a larger size leads to a more significant improvement in downstream task performance. To boost the performance while reining in communication overhead, we propose to allocate fine-tuning parameters to individual blocks according to their importance levels under a constraint on the total parameter size. In contrast, in current techniques, the number of such parameters is kept uniform across blocks. The design is inspired by the observation that the model weights at different locations in a FoMo show different importance levels for fine-tuning. Considering the low-rank reparameterzation method, the importance of a fine-tuning matrix in a transformer can be measured by the magnitude of singular values of the matrix. Alternatively, the ablation method can be applied to evaluate the important of a fine-tuning component by measuring the training loss increment due to the removal of the component, which can be approximated by the gradient-weight product [11]. Considering parameter-representation based PEFT, the parameter allocation is reflected in controlling the ranks of low-rank reparameterization in different blocks. In the case of Prefix-tuning, the allocation involves controlling the prefix sizes of individual blocks.

## 4.2 Over-the-Air Computing for DEFT

For DEFT implementing adapter-based fine-tuning, the adapter, which is a lightweight neural network, can be kept and updated on the device while the remaining FoMo executed at the server. The high-dimensional message exchange between the two network components places a heavy burden on the air interface. We propose to lift the burden by applying AirComp to realize neural connections over the air. Traditional AirComp techniques exploit the radio waves 'superposition principle to realize over-the-air signal aggregation in a distributed computing system, thereby solving the scalability problem in multi-access [12]. The novelty of the proposed design lies in the over-the-air realization of matrix-vector multiplication, a common operation in neural connections. Thereby, turning interference into a computation mechanism avoids the need to transmit neuron outputs using many orthogonal channels. The specific design leverages precoding and post-equalization to convert a multi-input-multi-output (MIMO) channel into a desired weight matrix. Thereby, the transmission of an analog modulated signal vector results in the receiver directly receiving the matrix-vector product result. It is also possible for MIMO AirComp to support over-the-air non-linear activation components in MLP layers such as ReLU and Sigmoid. One approach is to approximate the activation functions into a set of linear AirComputable sub-functions. An alternative one can use a non-linear detector at the receiver, whose parameters can be calibrated according to the characteristics of activation layers.

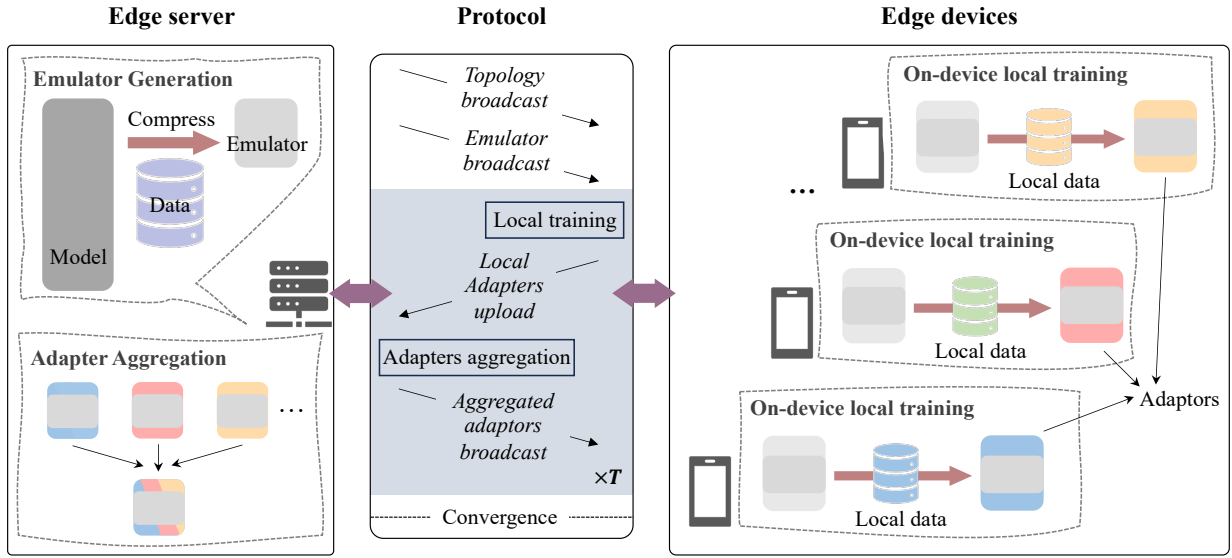

Fig. 3. Federated device-edge fine-tuning.

# 5. Multi-User DEFT

Involving multiple users to cooperate in fine-tuning has several advantages. First, leveraging training data distributed at users helps to meet the need for more data as required by fast few-shot fine-tuning. Second, the cooperation makes DEFT more efficient as a FoMo can be simultaneously fine-tuned to multiple downstream tasks. Last, the DEFT process can exploit distributed computation resources. These motivate us to propose three relevant techniques, namely federated DEFT, distributed prompt ensemble, and prompt transfer, in the following sub-sections.

## 5.1 Federated DEFT

The federated DEFT coordinates multiple users with similar tasks to fine-tune a FoMo that serves all users simultaneously. The technique exploits distributed data while preserving their ownership by avoiding raw data uploading. This is realized by applying the well-known Federated Edge Learning (FEEL) that iteratively trains a global model at the server by efficiently utilizing personalized data at devices. Specifically, in each iteration, the edge devices compute gradient (or sub-models) based on their datasets and the server aggregates these immediate results to update (or assemble) a global model. This process is repeated until the global model converges. Building on FEEL, two federated DEFT paradigms are presented as follows.

*1) Emulating FoMo for Local Fine-tuning:* A FoMo is needed to compute gradients for updating fine-tuning parameters (e.g., adapter or prompt encoder) locally at a device but downloading the FoMo is infeasible as mentioned. A natural approach is to download a compressed version to act as an emulator of the original model. Researchers have developed sophisticated techniques for emulator compression that adopt knowledge distillation to drop transformer layers while making an attempt to maximize the approximation accuracy [13]. The federated learning framework of emulating FoMo is depicted in Fig. 3. Provisioned with an emulator, a device is then able to compute a stochastic gradient using its local dataset. However, the gradients deviate from the optimal one due to limited local data. This makes it necessary to aggregate local gradients across devices using the FEEL technique. Specifically, in each iteration of federated DEFT, all devices upload local gradients to a server for aggregation, which ten broadcasts the aggregated gradient to devices for updating fine-tuning parameters. The avoidance of uploading data embeddings gives the emulator-based technique the advantages of privacy preservation and relatively low communication overhead. FoMo emulation is still at its nascent stage and the state-of-the-art techniques can achieve a reduced model size to a fraction, say, 20%. This limits the application of emulator-based federated DEFT to small-scale FoMo (e.g., no more than 1 billion parameters) and high-end smartphones. This calls for the development of more efficient FoMo compression techniques.

2) *Server FoMo Assisted Local Fine-tuning:* This federated DEFT technique targets two PEFT techniques, P-tuning, and adapters (which are resided in the first few layers of a fine-tuned FoMo). Its essential idea is for the FoMo at a server to compute and provide a gradient to all devices for local updating of fine-tuning parameters, i.e., prompt encoder in the case of P-tuning or adapters. Consider P-tuning and a single iteration of federated DEFT has the following procedure. Specifically, a tunable prompt encoder is employed at devices to generate intermediate soft prompts that are appended to data embeddings [5]. The results are then efficiently uploaded (using e.g., AirComp) to the server for aggregation before input into the FoMo for aggregated gradient computation via back-propagation. In the case of the adaptor-based tuning, local adapter outputs, which result from input of local data embeddings, are uploaded and then aggregated instead. For other PEFT techniques, the implementation of the above federated DEFT techniques is possible in principle but more tedious as the fine-tuning parameters are distributed over different blocks of the FoMo [see prefix-tuning or parameter representation in Fig. 1(b)]. This calls for new intelligent designs.

## 5.2 On-the-Fly Prompt Ensemble Boosting

Recent research findings show a prompt ensemble, which comprises a set of lightweight prompts, outperforms a single prompt [14]. This inspires us to propose the technique of on-the-fly boosting (FlyBoosting) to progressively broaden the range of solvable problems of a prompt ensemble, thereby realizing communication-and-computation efficient multi-device DEFT. Starting with an initial ensemble (prompt set) at the server, each iteration of FlyBoosting, which is based on a well-known approach called *chain of thought*, consists of three steps. First, each device tests the effectiveness of the current ensemble for the local task by asking the server to use it together with the FoMo to fill in the blanks in a set of reasoning paths and compares the answers with their group truth known locally. A simple example of a reasoning path is "Neymar plays for team ___; The team plays in city ___." This allows each device to generate a local-effectiveness score for the ensemble. Second, a server selects those devices with low scores to generate new prompts, which better match their local task, for modifying the ensemble (element addition or replacement). As a result, the problem-solving range of the current ensemble is enhanced. FlyBoosting is communication efficient as it involves only a subset of devices to update the ensemble in each round. The above steps are repeated until all devices are satisfied with the ensemble with, e.g., their local-effectiveness scores exceeding a given threshold. Among others, designing goal-oriented communication techniques for FlyBoosting is a promising direction. One particular opportunity is importance-aware scheduling and resource allocation where the importance of a device is inversely proportional to its local-effectiveness score. The rationale is that a device with a lower score contributes more significantly to enhancing the ensemble's capability.

The scheduler should balance the score and channel condition by using a metric combining the score with channel gain/rate.

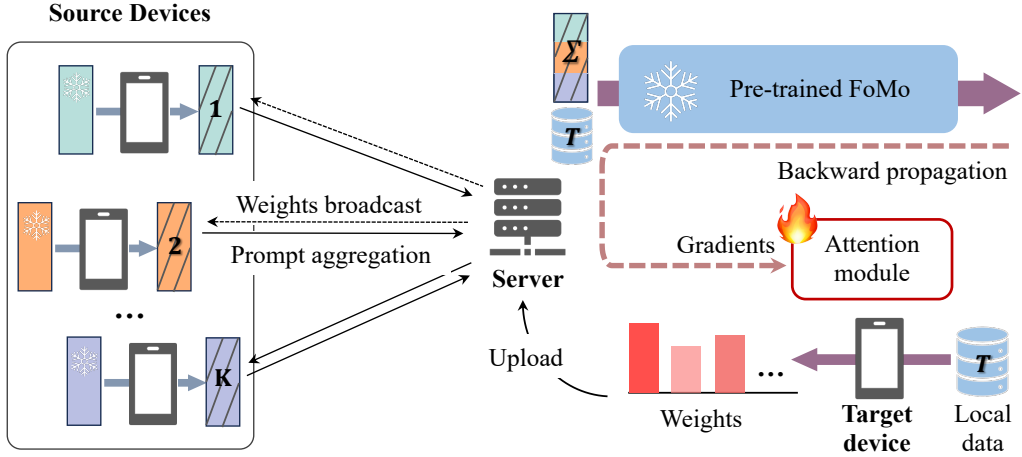

Fig. 4. D2D prompt transfer-and-fusion

## 5.3 Device-to-Device (D2D) Prompt Transfer-and-Fusion

Consider a cluster of devices with pre-trained prompts (or prompt encoder models) and the admission into the cluster a new device, called target device with task relevance or similarity to some existing devices, called source devices. We propose D2D prompt transfer, a form of knowledge transfer, from source devices to the target device to reduce fine-tuning overhead. The system operations are illustrated in Fig. 4 and explained as follows. The underpinning method is interpolation of source prompts using a trainable attention modular to learn on measuring the contribution of each source model [15]. Interpolation has demonstrated effective even via simple model fusion, namely that the desired prediction in the target task can be performed by a combination of outputs of different prompt-encoder models at source devices. Prompt transfer leveraging peers' assistance is faster and more efficient than fine-tuning from scratch. It can be implemented either in a distributed network with D2D links or a network with server coordination. From the perspective of air interface, the prompt interpolation and model fusion are essentially linear operations (e.g., weighted combination of source prompt matrices) and thus can be efficiently realized over the air using AirComp.

# 6. Experiments

## 6.1 Experiment settings

We consider the employment of P-tuning techniques (§ 2.1) to fine-tune two prevalent pre-trained FoMos: 1) Megatron-11B with 11 billion parameters and 2) T5-base with 220 million parameters. To this end, a server with Nvidia A100 GPUs is employed. The wireless link connecting the device-server pair is modeled as a single-input-single-output Gaussian channel with a bandwidth of 20MHz. We consider both digital transmission and (analog) AirComp. Consider the digital-transmission case. Each transmitted coefficient is quantized into 8 bits and the communication rate set as the channel capacity that depends on the signal-to-noise-ratio (SNR). Moreover, multi-access is based on frequency division orthogonal access. For the case of AirComp, the symbol rate is 1/20MHz with each symbol modulated with one real coefficient using uncoded linear analog modulation. The settings for single-device and multi-device DEFT are as follows.

*1) Single-device DEFT:* The target downstream task is knowledge probing on the popular dataset of LAMA-29k, where the fine-tuned Megatron-11B is required to perform a cloze test that involves filling a portion of masked text (see [14] for more details). The number of epochs is set as 5 by default, each of which comprises 91 iterations. The fine-tuning parameters for Megatron-11B are those of a on-device prompt encoder that comprises around 132 million weights. The prompt encoder and dataset are owned by the edge device whose computation capability is set as 1/50 of the server's.

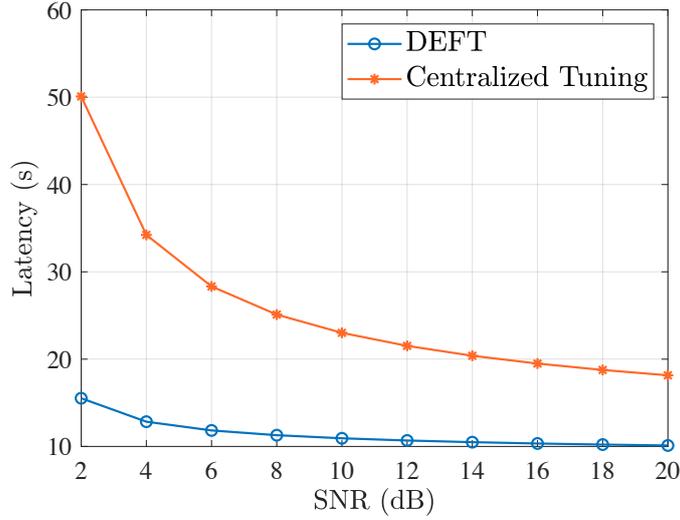

Fig. 5. Latency performance comparison between DEFT and centralized fine-tuning.

*2) Multi-device DEFT:* We consider D2D prompt transfer (§5.3) to obtain a target soft prompt consisting of about 2 million parameters. The target prompt is appended to a pre-trained T5-base to perform the well-known downstream task of GLUE (General Language Understanding Evaluation) (see [15] for more details). Each epoch comprises 115 iterations. The prompt fusion module (i.e., attention module) is operated by the server. In each iteration, for the purpose of prompt transfer, each source device uploads the element-wise product of their fixed prompts and the attention matrix broadcasted by the server.

## 6.2. Performance of Single-device DEFT

We compare the performance of the proposed single-user DEFT with the centralized (server) fine-tuning. The former requires the device uploads its prompt embeddings in each iteration. On the contrary, the latter, a benchmarking scheme, requires the device to upload the raw data, i.e., LAMA-29K. Reliable digital transmission is used for both schemes and thus the task-precision performance of their corresponding fine-tuned FoMos is identical. However, DEFT outperforms centralized tuning in terms of (end-to-end) latency. The latency consists of both computation and communication delays accumulated over the whole fine-tuning process. In Fig. 5, the curves of latency required versus SNR are plotted for both DEFT and centralized fine-tuning. From Fig. 5, DEFT is shown to substantially reduce the latency as opposed to the centralized tuning as prompt embeddings are much smaller in size than high-dimensional raw data. In particular, 5-time latency reduction is observed at SNR of 10 dB. The performance gap narrows as the SNR increases. The resultant higher communication rate helps centralized tuning significantly as communication latency dominates its computation counterpart due to a very powerful A100 server. In contrast, DEFT is less sensitive to the rate changes since local computation for fine-tuning constitutes a bottleneck especially in the high SNR region. This is reflected in the saturation of the latency curve in the region.

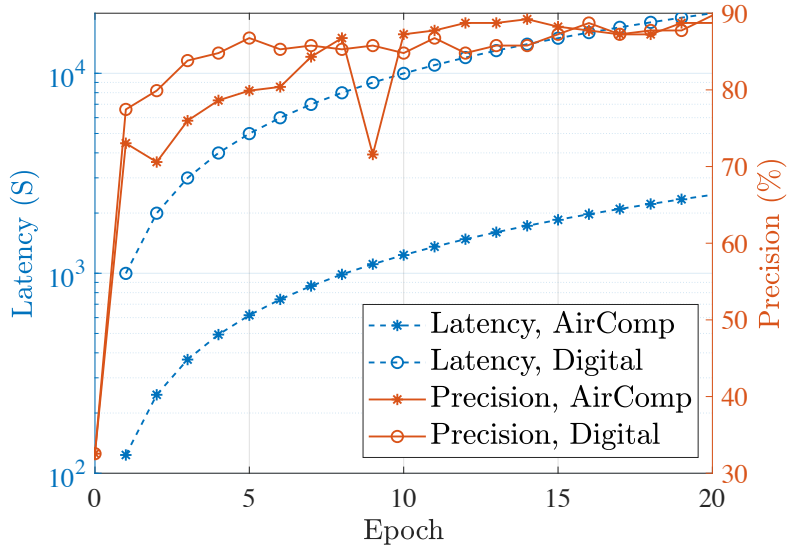

Fig. 6. Comparison of latency and task precision between DEFT using AirComp and digital transmission.

### 6.3. Performance of Multi-device DEFT

Multi-device DEFT employs AirComp to realize ultra-fast over-the-air fusion of attention weighted prompts from source devices. It is benchmarked against the traditional scheme of frequency division orthogonal access for digital transmission. The curves of task precision (i.e., averaged task execution accuracy on the evaluation set) and communication latency (accumulated over the iterative fine-tuning process) and versus the number of fine-tuning epochs are plotted in Fig. 6 in solid and dashed lines, respectively. The SNR is set as 20dB.

*1) Task precision of fine-tuned FoMo:* For both schemes in comparison, the task precision is observed from Fig. 6 to rapidly increase in the first five epochs and then exhibit a plateau in the following epochs. For AirComp without coding, the exposure of transmitted prompts to channel noise degrades the precision with respect to its digital counterpart. The performance gap, however, narrows as the training progresses. Specifically, the precision of both schemes simultaneously reachs 90% after 20 epochs. This shows the iterative fine-tuning process together with multi-device averaging helps to rein in the effect of channel distortion caused by AirComp without coding.

*2) Communication latency performance:* For both schemes, the communication latency is incurred by the transmission of outputs of prompt encoder, which is the component being fine-tuned, and summed over the number of epochs. One can see in Fig. 6 that AirComp-assisted DEFT is much faster than DEFT based on orthogonal access, e.g., achieving around 10-time latency reduction at 15 epochs. The precision-and-latency results advocate AirComp as an ultra-fast air interface for DEFT based on D2D prompt transfer-and-fusion.

# 7. Concluding Remarks

With their human-like capabilities, FoMos (or generative AI models) are expected to revolutionize fields in engineering and science. In the context of 6G, we will see wide development of FoMos as a platform to automate next-generation mobile tasks. Researchers believe that scaling up of AI is reaching the peak and the next phase shall focus on fine-tuning FoMos to downstream tasks. Then the proposed area of DEFT presents a goldmine of research opportunities in communication-computation-integrated designs. Let us conclude this article by describing several promising opportunities.

*1) Communication and computation balancing:* While keeping fine-tuning components on the device, DEFT can download some frozen layers of FoMos to reduce communication overhead and to improve privacy at the cost of higher local computation load. This necessitates the optimization of FoMo partitioning and downloading of FoMos balance communication and computation.

*2) Distributed computing for collective intelligence:* FoMos can solve a complicated task by breaking down higher-level prompts into lower-level sub-tasks for distribution to connected devices. The task splitting and distribution need to be optimized based on the principle of matching the heterogeneous devices 'capabilities, i.e., computation power, communication rate, sensors and actuators, to sub-tasks' requirements.

*3) Resource management for dynamic DEFT:* The on-device FoMo components need to be updated periodically for adaptation time tasks. This calls for developing online DEFT including queue management, predictive device clustering, and dynamic resource allocation in the process of cooperative tuning.

*4) DEFT for hierarchical FoMos:* In mobile networks, FoMos are envisioned to be implemented hierarchically with components offloaded to edge devices, servers, and the cloud. Then, DEFT needs to be extended to support the architecture where fine-tuning of FoMos can occur flexibly at an arbitrary networking layer for layer-specified goals. For instance, on-device fine-tuning targets personalized services, while on-server tuning adapts to common requirements of users under its coverage based on behavior analysis.